\documentstyle[aps,prd,eqsecnum,epsf]{revtex}

\begin{document}

\newcommand{\gsim}{\mbox{\raisebox{-1.ex}{$\stackrel
     {\textstyle>}{\textstyle\sim}$}}} 
\newcommand{\lsim}{\mbox{\raisebox{-1.ex}{$\stackrel
     {\textstyle<}{\textstyle \sim}$}}} 
\newcommand{\square}{\kern1pt\vbox{\hrule height
1.2pt\hbox{\vrule width 1.2pt\hskip 3pt
   \vbox{\vskip 6pt}\hskip 3pt\vrule width 0.6pt}\hrule
height 0.6pt}\kern1pt}

\newcommand{\singlefig}[2]{
\begin{center}
\begin{minipage}{#1}
\epsfxsize=#1
\epsffile{#2}
\end{minipage}
\end{center}}
%
\newenvironment{figcaption}[2]{
 \vspace{0.3cm}
 \refstepcounter{figure}
 \label{#1}
 \begin{center}
 \begin{minipage}{#2}
 \begingroup \small FIG. \thefigure: }{
 \endgroup
 \end{minipage}
 \end{center}}
%


\draft
\title{Preheating with extra dimensions}
\author{Shinji Tsujikawa \thanks{electronic
address:shinji@gravity.phys.waseda.ac.jp}}
\address{ Department of Physics, Waseda University,
3-4-1 Ohkubo, Shinjuku-ku, Tokyo 169-8555, Japan\\[.3em]}
\date{\today}
\maketitle
\begin{abstract}
We investigate preheating in a higher-dimensional 
generalized Kaluza-Klein theory with a quadratic 
inflaton potential $V(\phi)=\frac12 m^2\phi^2$ including 
metric perturbations explicitly.
The system we consider is the multi-field model where
there exists a dilaton field $\sigma$ which corresponds 
to the scale of compactifications and another scalar field
$\chi$ coupled to inflaton with the interaction
$\frac12 g^2\phi^2\chi^2+\tilde{g}^2\phi^3\chi$.
In the case of $\tilde{g}=0$, we find that the perturbation
of dilaton does not undergo parametric amplification 
while the $\chi$ field fluctuation can be enhanced 
in the usual manner by parametric resonance.
In the presence of the $\tilde{g}^2\phi^3\chi$ coupling,
the dilaton fluctuation in sub-Hubble 
scales is modestly amplified by the growth of 
metric perturbations for the large coupling $\tilde{g}$. 
In super-Hubble scales, 
the enhancement of the dilaton fluctuation 
as well as metric perturbations is weak, taking into account
the backreaction effect of created $\chi$ particles.
We argue that not only is it possible to predict the 
ordinary inflationary spectrum in large scales
but extra dimensions can be
held static during preheating in our scenario. 
\end{abstract}


\baselineskip = 18pt

%
\section{Introduction}                            %
Physics in higher-dimensions has received
much attention in an attempt to unify the interactions in Nature
originating from Kaluza and Klein.
For example, the ten-dimensional 
$E_8 \times E_8$ heterotic superstring theory 
would be a strong candidate to describe the real world\cite{GSW}.
The strong coupling limit of this theory
was found to be the eleven-dimensional supergravity\cite{HW},
which is equivalent to the low-energy effective M-theory. 
Since the spacetime we recognize is four-dimensional, 
we conventionally utilize some mechanisms of
dimensional reduction assuming that extra dimensions 
are compactified on some manifolds (Kaluza-Klein reductions).
Recently, Randall and Sundrum\cite{brane1} proposed an alternative way 
of compactifications based on brane models, which was 
originally introduced as a solution to the 
hierarchy problem between the weak and 
Planck scale\cite{brane2,brane3}. 
In the brane scenario, gravity works in the five-dimensional
bulk of spacetime while matter fields are confined
in four-dimensions. 
In this paper, however, we adopt conventional Kaluza-Klein reduction
from a higher-dimensional action whose dimension is larger than five.

One of the most important topics which plagues
such higher-dimensional theories is the stability 
of extra dimensions.
The internal dimensions
need to be held static against small fluctuations
in order to settle in the present universe which would be
a direct product of the four-dimensional Minkowski spacetime 
$M^4$ and a small internal space $K^d$. 
In this respect, the basic argument is to introduce a cosmological constant
in the higher-dimensional action and keep extra dimensions static
with the existence of some fields.
In such models  including the Candelas-Weinberg (CW) 
model\cite{CW} (sphere compactification with a 
cosmological constant and the one-loop
quantum correction), it has been recognized that 
the present vacuum is static against linear perturbations\cite{BLV}
and even non-linear large perturbations\cite{maeda1}.
It was also found that stability of the internal space is preserved
against a quantum tunneling without the topology 
change\cite{maeda2}.

From a cosmological point of view, it is natural to
ask whether the internal space is held static during 
an inflationary epoch\cite{inflation}.
Amendola {\it et al.}\cite{stabilityinf}
considered stability of 
compactification in the CW model
with the existence of an inflaton field $\phi$.
In old, new, and extended inflation models, the 
system exhibits semiclassical properties in
which the stability is preserved as long as the transition
probability for the scale of the internal space to tunnel 
through its potential barrier is smaller than that of inflaton.
In the chaotic inflation model, the expansion of the internal 
space can be classically avoided if we choose 
the initial value of inflaton and parameters 
of the model appropriately. 
Amendola {\it et al.} also obtained the upper bound 
for the present scale of the internal space as 
$b_*~\lsim~\sqrt{d} \times 10^5 l_{\rm pl}$,
where $l_{\rm pl}$ is the Planck length, 
by the requirement of successful chaotic
inflation and stability of compactifications.
As for the inflationary scenario in generalized 
Einstein theories including Brans-Dicke and induced gravity
theories, Berkin and Maeda\cite{BM} analyzed the 
dynamics of inflation in new and chaotic inflationary 
models in the presence of
a dilaton field $\sigma$ with a potential $U(\sigma)=0$. 
In the context of large internal dimensions, several 
authors\cite{LI} recently investigated inflation 
with the higher-dimensional fundamental 
Planck scale in the TeV region. 

After the inflationary period ends, the system enters
a reheating stage. 
It has been recognized that reheating will turn on
by an explosive particle production which is called 
{\it preheating}\cite{pre1,KLS1,KLS2}.
As compared with other inflationary models, the chaotic 
inflationary scenario typically leads to 
the strong amplification of a scalar field $\chi$ 
coupled to inflaton with the interaction
$\frac12 g^2\phi^2\chi^2$ due to parametric resonance
during the oscillating stage of inflaton.
In this scenario, a lot of works have already been done
using analytic approaches based on the Mathieu\cite{KLS1,KLS2} 
or Lam\'e equations\cite{structure} and numerical computations
by mean field approximations\cite{KTmassivehartree,Boy,Baa} 
or full lattice simulations\cite{KTselffull,KTmassivefull,PR}.
The existence of the preheating stage is important in the 
sense that it would affect the 
baryogenesis in grand unified scale\cite{baryogenesis}, 
topological defect formation\cite{defect}, nonthermal 
phase transition\cite{phasetransition}, and 
gravitational waves\cite{GW}.

Recently, Mazumdar and Mendes\cite{MM} considered 
preheating in generalized Einstein theories including 
higher-dimensional theories in the massive chaotic 
inflationary model $V(\phi)=\frac12 m^2\phi^2$ 
with a scalar field $\chi$ coupled to inflaton. 
After dimensional reductions,  
we have a dilaton field $\sigma$ which corresponds to 
the radius of extra dimensions.
They investigated the multi-field system of scalar fields
$\phi$, $\chi$, $\sigma$ in the case 
where dilaton does not have its own potential.
It was pointed out that long-wave modes ($k \to 0$)
of the fluctuation of dilaton can be enhanced 
due to the growth of metric perturbations.
Although their scheme is based on the torus
compactifications which have zero curvature,
compactifications on the sphere give rise to 
a potential term due to the curvature of the internal 
space. It is worth investigating whether the stability of 
compactification is preserved or not during preheating, 
when dilaton has its own potential.
In this paper, we adopt the dilaton potential  
based on the CW model in the presence of 
a massive inflaton field,
and analyze the evolution of scalar field fluctuations
including the backreaction effect of created particles.
We include metric perturbations explicitly for
the evolution equations, and also investigate whether 
this model predicts the density perturbation observed by
the Cosmic Background Explorer (COBE) satellite.

This paper is organized as follows.
In the next section, we describe our model and 
consider the dynamics of inflation in the presence
of the dilaton field.
In Sec.~III, we investigate the parametric amplification
of field fluctuations during preheating 
including metric perturbations.
It is particular of interest to study the evolutions of super-Hubble
dilaton and metric perturbations.
We present our conclusion and discussion in the final section.

\section{Inflation with extra dimensions}   

We investigate a model in $D=d+4$ dimensions
with a cosmological constant $\bar{\Lambda}$ and 
a single scalar field $\bar{\phi}$, 
\begin{eqnarray}
S=\int d^D x \sqrt{-\bar{g}} \left[ \frac{1}
{2\bar{\kappa}^2} \bar{R}-2\bar{\Lambda}
+\bar{{\cal L}}(\bar{\phi}) \right],
\label{B1}
\end{eqnarray}
where $\bar{\kappa}^{2}/8\pi \equiv 
\bar{G}$ and $\bar{R}$
are the $D$-dimensional gravitational constant 
and a scalar curvature 
with respect to the $D$-dimensional metric
$\bar{g}_{MN}$, respectively.
The Lagrangian density for the minimally coupled 
$\bar{\phi}$ field is written as 
\begin{eqnarray}
\bar{{\cal L}}(\bar{\phi})=
-\frac12 \bar{g}^{MN}\partial_M \bar{\phi}
\partial_N \bar{\phi}-\bar{V}(\bar{\phi}),
\label{B2}
\end{eqnarray}
where $\bar{V}(\bar{\phi})$ is a potential of the
$\bar{\phi}$ field.

We compactify extra dimensions to a $d$-dimensional 
sphere $S^d$. Then the metric 
$\bar{g}_{MN}$ is expressed as
\begin{eqnarray}
ds^2_D=\bar{g}_{MN}dx^M dx^N
=\hat{g}_{\mu\nu}dx^{\mu}dx^{\nu}
+b^2 ds^2_d,
\label{B3}
\end{eqnarray}
where $\hat{g}_{\mu\nu}$ is a four-dimensional metric,
$b$ is a scale of the $d$-dimensional sphere
whose present value is $b_*$, and 
$ds^2_d$ is a line element of the $d$-unit sphere.
After dimensional reduction,
the action $(\ref{B1})$ yields
\begin{eqnarray}
S=\int d^4 x \sqrt{-\hat{g}} \left(\frac{b}{b_*}\right)^d
 \Biggl[\frac{1}{2\kappa^2} \Biggl\{
\hat{R} +d(d-1) \frac{\partial_{\mu}b \partial_{\nu}b}{b^2}
\hat{g}^{\mu\nu}
+\frac{d(d-1)}{b^2} \Biggr\}
+V_d^0 \left\{ \hat{{\cal L}} (\hat{\phi})
-2\bar{\Lambda} \right\} \Biggr],
\label{B4}
\end{eqnarray}
where $\kappa^{2}/8\pi$
is Newton's gravitational constant which is expressed as
$\kappa^{2}/8\pi=\bar{\kappa}^{2}/(8\pi V_d^0)$
with the present volume of the internal space $V_d^0$,
and $\hat{R}$ is a scalar curvature with respect to
$\hat{g}_{\mu\nu}$.
The action $(\ref{B4})$ is different from the form of 
the Einstein-Hilbert action due to the 
time-dependent term $(b/b_*)^d$.
In order to obtain the usual form, we perform 
the following conformal transformation,
\begin{eqnarray}
\hat{g}_{\mu\nu}=\exp\left(-d\frac{\sigma}{\sigma_*}
\right) g_{\mu\nu},
\label{B5}
\end{eqnarray}
where $\sigma$ is the so-called dilaton field
which is defined by  
\begin{eqnarray}
\sigma &=& \sigma_* {\rm ln} \left( \frac{b}{b_*} \right), \\
\sigma_* &=& \left[ \frac{d(d+2)}{2\kappa^2} \right]^{1/2}.
\label{B7}
\end{eqnarray}
Then the four-dimensional action in the Einstein frame
can be described as
\begin{eqnarray}
S=\int d^4 x \sqrt{-g} \left[ \frac{1}{2\kappa^2}R
-\frac12 g^{\mu\nu} \partial_{\mu} \sigma 
\partial_{\nu} \sigma -U(\sigma)
-\frac12 g^{\mu\nu} \partial_{\mu} \phi 
\partial_{\nu} \phi -\exp \left(-d \frac{\sigma}{\sigma_*}
\right) V(\phi) \right],
\label{B8}
\end{eqnarray}
where $R$ is a scalar curvature related with $g_{\mu\nu}$,
and a scalar field $\phi$ is defined by $\phi=\sqrt{V^0_d}
\hat{\phi}$.
The potential $U(\sigma)$ for the $\sigma$ field is
expressed as
\begin{eqnarray}
U(\sigma)=\frac{\bar{\Lambda}}{\kappa^2}
e^{-d\sigma/\sigma_*}-
\frac{d(d-1)}{2\kappa^2b_*^2}
e^{-(d+2)\sigma/\sigma_*}.
\label{B9}
\end{eqnarray}
The second term in Eq.~$(\ref{B9})$ appears due to 
the curvature of the internal space by compactifications
on the sphere $S^d$.
However, since the potential $(\ref{B9})$ lacks a local
minimum to stabilize the dilaton field, one needs to introduce
quantum correction effects which are so-called Casimir effects. 

Adding a one-loop effective action which is proportional
to $e^{-2(d+2)\sigma/\sigma_*}$
to the potential $U(\sigma)$
and imposing the conditions that the $\sigma$ field has 
a local minimum at $\sigma=0$ and its extremum is zero,
$U(\sigma)$ can be written in the following form\cite{mukoh} :
\begin{eqnarray}
U(\sigma)=\alpha \left[ \frac{2}{d+2}
e^{-2(d+2)\sigma/\sigma_*}+e^{-d\sigma/\sigma_*}
-\frac{d+4}{d+2} e^{-(d+2)\sigma/\sigma_*} \right],
\label{B10}
\end{eqnarray}
with
\begin{eqnarray}
\alpha=\frac{d(d-1)(d+2)}
{2\kappa^2(d+4)b_*^2}.
\label{B11}
\end{eqnarray}
The first, second, and third terms in $(\ref{B10})$ 
are due to the Casimir energy, the cosmological 
constant, and the curvature of the internal space, respectively.
The potential $U(\sigma)$ in the action $(\ref{B8})$ has 
a local minimum at $\sigma=0$ and 
a local maximum at $\sigma_c (>0)$ which depends on
the extra dimension $d$.
In order to reach the final state $\sigma=0$ which 
corresponds to the present scale of the internal space
$b=b_*$, the initial value of $\sigma$ is required 
to be $0<\sigma_I<\sigma_c$ (we assume $\sigma_I>0$),
where the subscript $I$ denotes the initial value.
Then $\sigma$ evolves toward the minimum of its 
potential, and begins to oscillate around $\sigma=0$.
Since extra dimensions are compactified on the sphere, 
this gives rise to the four-dimensional Kaluza-Klein
field $\psi_{lm}$ whose mass is given by
$M_l^2=l(l+d-1)e^{-(d+2)\sigma/\sigma_*}/b_*^2 $.
It was suggested in Ref.~\cite{mukoh} that 
Kaluza-Klein modes can be
excited by the oscillation of the $\sigma$ field 
in the flat Friedmann-Robertson-Walker (FRW)
background. Later, we found that catastrophic enhancement
of Kaluza-Klein modes does not occur relevantly 
for any values of 
$\sigma_I$ and the quantum number $l \ge 1$\cite{shinji}.
Hence we only consider the case of $l=0$ in this paper.
For a complete study, however, we have to take into account
the existence of Kaluza-Klein modes with $l \ge 1$.

In the presence of the inflaton field $\phi$, the effective 
potential for the dilaton field is described by the action 
$(\ref{B8})$ as follows
\begin{eqnarray}
U_1(\sigma,\phi)=\alpha \left[ \frac{2}{d+2}
e^{-2(d+2)\sigma/\sigma_*}+
e^{-d\sigma/\sigma_*} \left\{1+\frac{V(\phi)}
{\alpha} \right\}
-\frac{d+4}{d+2} e^{-(d+2)\sigma/\sigma_*} \right].
\label{B12}
\end{eqnarray}
The stability of compactification during inflation 
with a potential $(\ref{B12})$ in several models of inflation 
was analyzed in Ref.~\cite{stabilityinf}.
Since we are interested in the model 
where strong parametric amplification of scalar fields 
can be expected during preheating, 
we adopt the quadratic potential of 
chaotic inflation,\footnote{In the model of the self-coupling potential
$V(\phi)=\frac14 \lambda\phi^4$, we will give some discussions
in the final section.}
\begin{eqnarray}
V(\phi)=\frac12 m^2\phi^2.
\label{B13}
\end{eqnarray}
In Fig.~1, we depict the effective potential $(\ref{B12})$
with $(\ref{B13})$.
As was pointed out in Ref.~\cite{stabilityinf},
$U_1(\sigma,\phi)$ has either two local extrema 
or no local extrema for a fixed value of $\phi$.
When $\phi$ is smaller than some critical value
$\phi_c$, the potential barrier which prevents 
the $\sigma$ field from going toward infinity exists,
and the scalar field evolves toward the potential minimum 
at $\phi=\sigma=0$.
However, when $\phi>\phi_c$, this barrier 
disappears and the internal space grows without 
limit. The critical value $\phi_c$ can be obtained 
by solving the equation $\partial U_1/\partial\sigma=
\partial^2 U_1/\partial\sigma^2=0$ as
\begin{eqnarray}
\phi_c^2=\frac{2\alpha}{m^2}
\left[\left(1+\frac{2}{d}\right)
\left(\frac12 \right)^{2/(d+2)} -1\right].
\label{B14}
\end{eqnarray}
In order to result in the present vacuum $\sigma=0$,
the inflaton $\phi$ is constrained as
\begin{eqnarray}
\phi^2 < \phi_c^2
=\frac{d(d-1)(d+2)}{8\pi(d+4)}
\left[\left(1+\frac{2}{d}\right)
\left(\frac12 \right)^{2/(d+2)} -1\right]
\left(\frac{m_{\rm pl}}{m}\right)^2
\frac{1}{b_*^2},
\label{B15}
\end{eqnarray}
where we have used Eq.~$(\ref{B11})$.
In the chaotic inflationary scenario, the initial value of 
inflaton is required to be $\phi_I~\gsim~3m_{\rm pl}$
in order to obtain the number of e-foldings greater than 60.
Further, the density perturbation observed by the COBE 
satellite constrains the coupling of inflaton as $m \sim 10^{-6}m_{\rm pl}$.
Then the condition $(\ref{B15})$ leads to the following
bound for the present value of the internal space:
\begin{eqnarray}
b_*^2~\lsim~\frac{d(d-1)(d+2)}{72\pi(d+4)}
\left[\left(1+\frac{2}{d}\right)
\left(\frac12 \right)^{2/(d+2)} -1\right]
\frac{10^{12}}{m_{\rm pl}^2}.
\label{B16}
\end{eqnarray}
For example, when $d=2$ and $d=6$, $b_*~\lsim~5\times
10^4/m_{\rm pl}$ and 
$b_*~\lsim~1.1 \times 10^5/m_{\rm pl}$, respectively.
For large values of $d$, Eq.~$(\ref{B16})$ reads
\begin{eqnarray}
b_*~\lsim~\sqrt{d} \times 10^5/m_{\rm pl}.
\label{B17}
\end{eqnarray}
As was suggested in Ref.~\cite{stabilityinf}, this value is
by about ten orders of magnitude smaller than the experimental
bound $b_*~\lsim~10^{17}/m_{\rm pl}$.
It is worth mentioning that such theoretical bound is derived 
by the analysis of stability of compactification
during inflation.
In what follows, we use the values of $b_*$
which satisfy the relation $(\ref{B16})$.

\begin{figure}
\begin{center}
\singlefig{12cm}{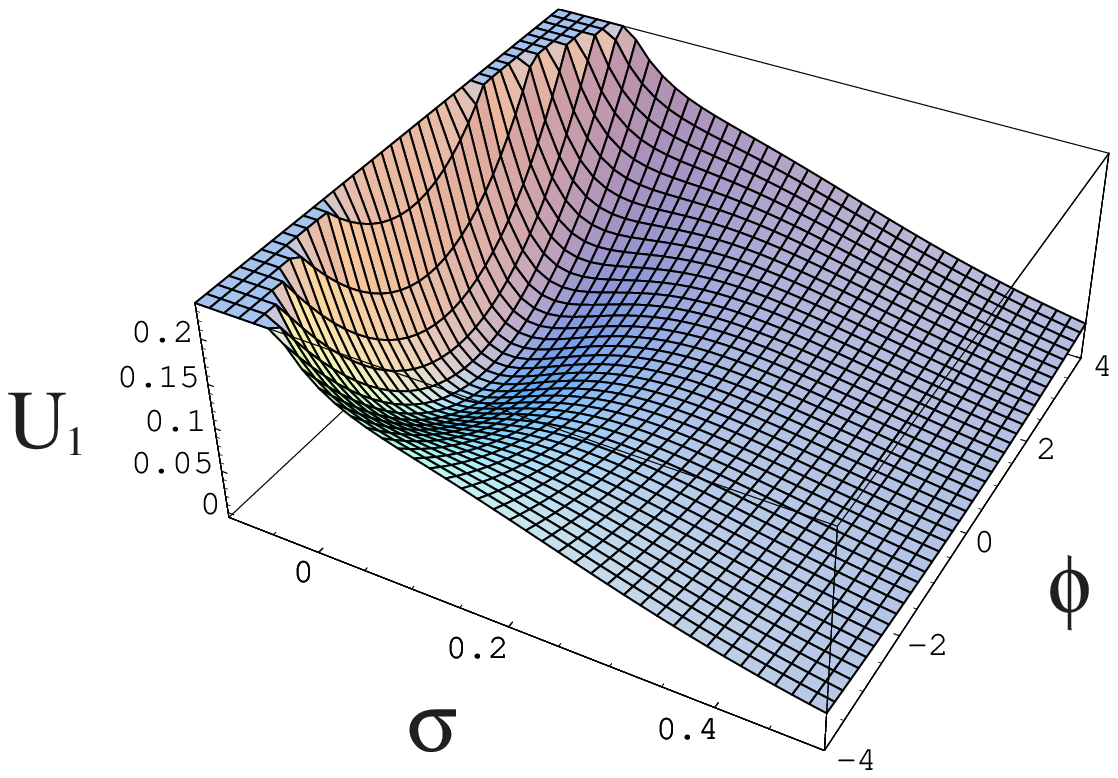}
\begin{figcaption}{Fig1}{12cm}
The effective potential $U_1(\sigma)$ which is obtained by 
introducing the Casimir energy in the sphere compactification
with $d=6$. We find that the potential barrier disappears for 
large values of inflaton. In order to evolve toward the present
vacuum state $\phi=\sigma=0$, inflaton is required to be 
smaller than some critical value $\phi_c$.
\end{figcaption}
\end{center}
\end{figure}

Let us consider the dynamics of inflation 
in the presence of the $\sigma$ field.
In the flat FRW background: $ds^2=-dt^2+a^2(t)d {\bf x^2}$, 
we find that the homogeneous parts of scalar 
fields and the scale factor satisfy the following
equations of motion by the action $(\ref{B8})$:
\begin{eqnarray}
\ddot{\phi}+3H \dot{\phi}+e^{-d\sigma/\sigma_*}
V'(\phi)=0,
\label{B19}
\end{eqnarray}
\begin{eqnarray}
\ddot{\sigma}+3H \dot{\sigma}+U'(\sigma)
-\frac{d}{\sigma_*} e^{-d\sigma/\sigma_*}
V(\phi)=0,
\label{B20}
\end{eqnarray}
\begin{eqnarray}
H^2 \equiv \left(\frac{\dot{a}}{a}\right)^2
     =\frac{\kappa^2}{3}
     \left[ \frac12 \dot{\phi}^2+e^{-d\sigma/\sigma_*}
     V(\phi)+\frac12 \dot{\sigma}^2+U(\sigma) \right],
\label{B21}
\end{eqnarray}
where $H$ is the Hubble expansion rate.

\begin{figure}
\begin{center}
\singlefig{12cm}{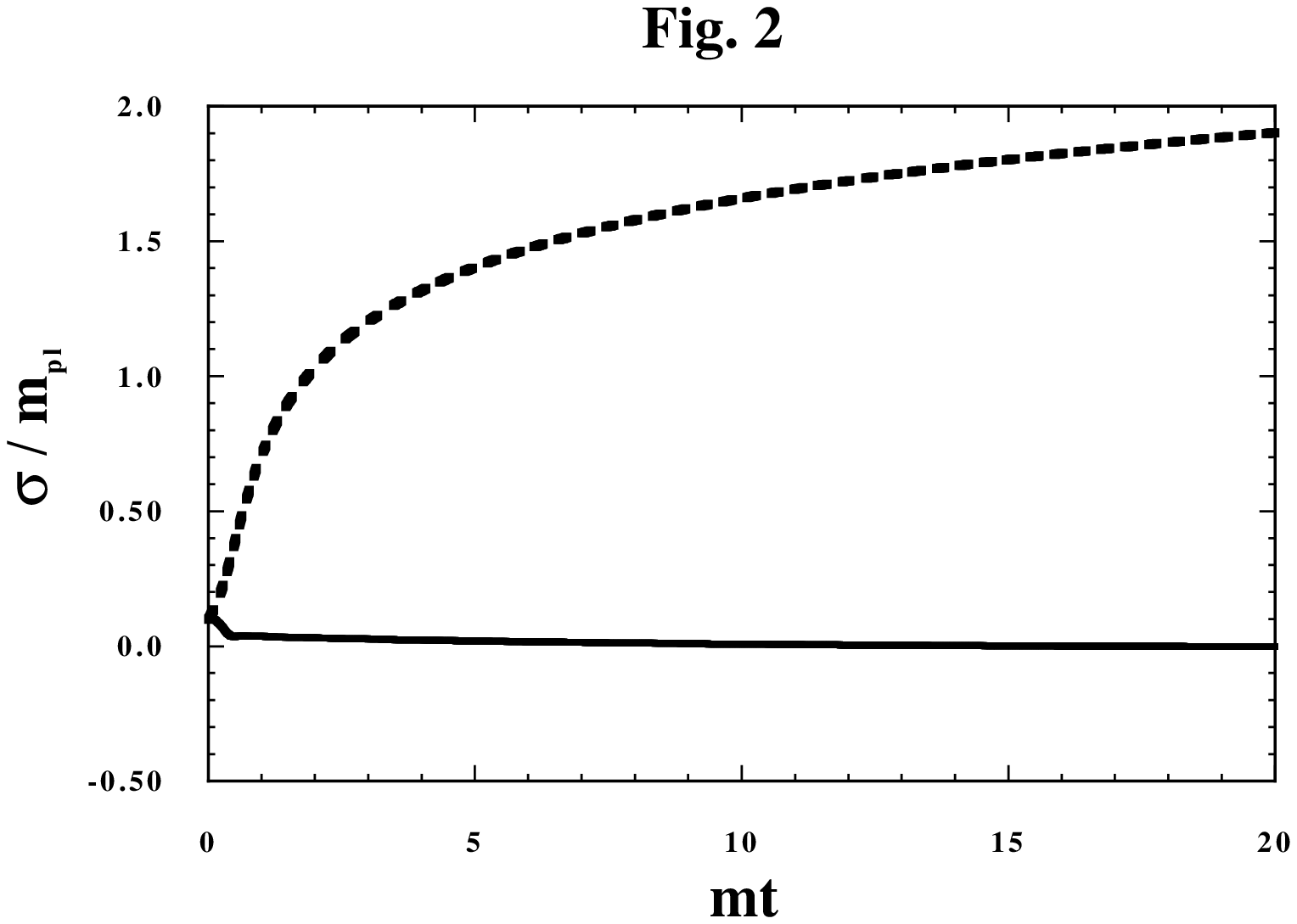}
\begin{figcaption}{Fig2}{12cm}
The evolution of the $\sigma$ field with the initial value of
$\sigma_I=0.1m_{\rm pl}$ for two cases of 
$\phi_I=3m_{\rm pl}$ (solid) and $\phi_I=4m_{\rm pl}$
(dotted) with $d=6$ and $b_*=1.0 \times 10^5/m_{\rm pl}$.
When $\phi_I=3m_{\rm pl}$, the $\sigma$ field evolves toward
the potential minimum at $\phi=\sigma=0$ due to the existence 
of the potential barrier. On the other hand, when 
$\phi_I=4m_{\rm pl}$, the internal space grows without limit. 
\end{figcaption}
\end{center} 
\end{figure}

When the initial value of inflaton is larger 
than $\phi_c$, the last term in the l.h.s. of 
Eq.~$(\ref{B20})$ dominates over the third 
term and the $\sigma$ field evolves
toward $\sigma=\infty$ (see Fig.~2).
This is the situation we want to avoid.
For the values of $\phi<\phi_c$, there exists 
a local minimum at $\sigma=\sigma_1$ and 
a local maximum at $\sigma=\sigma_2$
with $\sigma_2>\sigma_1>0$.
As long as the initial value of $\sigma$ exists
in the range of $\sigma<\sigma_2$, 
the $\sigma$ field evolves toward the potential
minimum at $\sigma=\sigma_1$.
The value of $\sigma_1$ decreases to zero as
the $\phi$ field moves toward the potential
minimum at $\phi=0$.
For the initial values of $\sigma$ and $\phi$
which are finally trapped in the potential minimum
at $\sigma=\phi=0$, one may consider that 
the dynamics of inflation is altered in the presence 
of the $\sigma$ field.
In this case, however, we can numerically confirm that
the third term in Eq.~$(\ref{B20})$ rapidly
makes the $\sigma $ field shift toward the local
minimum at $\sigma=\sigma_1$ for the values of
$b_*$ which satisfy the condition of Eq.~$(\ref{B16})$.
Then $\sigma$ begins to roll down along the valley
of $\sigma=\sigma_1$, and decreases as inflaton 
approaches its potential minimum. This behavior is found in 
Fig.~2. The dynamics of inflation is hardly affected by 
the presence of the $\sigma$ field, and the system can 
be effectively described by one scalar field $\phi$.

In Fig.~3, we plot the evolution of both the inflaton 
field and the number of e-foldings $N \equiv \ln (a/a_I)$
during inflation. After the rapid 
decrease of $\sigma$ at the initial stage, inflaton
slowly rolls down along its potential in the usual manner,
which results in sufficient inflation to solve several 
cosmological puzzles.
We also find that the number of e-foldings exceeds
$N \sim 60$ for the initial value of 
$\phi_I=3m_{\rm pl}$. The inflationary period ends 
when inflaton decreases to $\phi \approx 0.2m_{\rm pl}$,
which corresponds to the time $mt \approx 20$
in Fig.~3.
In the next section, we investigate the dynamics of field
and metric perturbations during preheating.

\begin{figure}
\begin{center}
\singlefig{12cm}{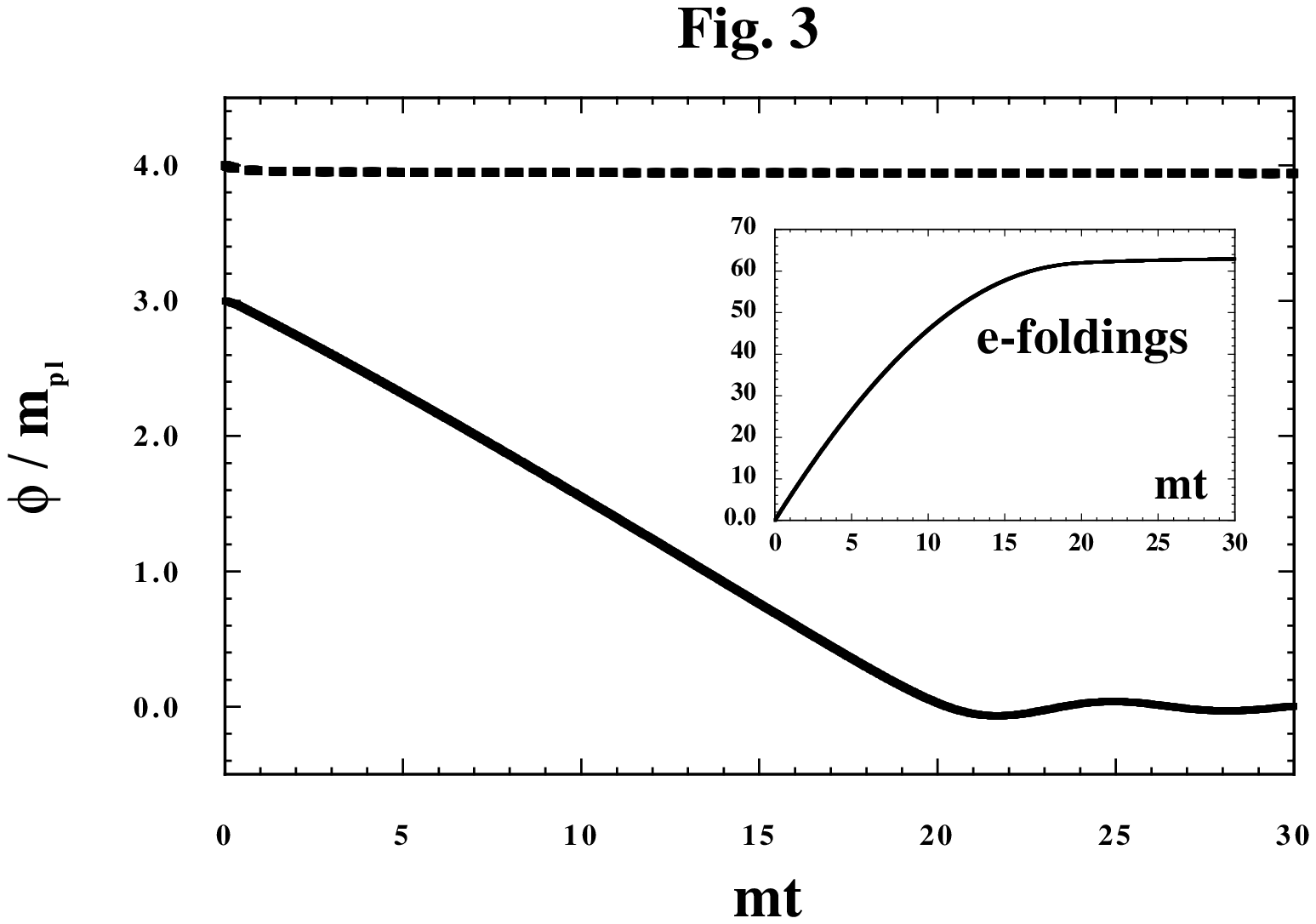}
\begin{figcaption}{Fig3}{12cm}
The evolution of the $\phi$ field during inflation 
with the initial value of
$\sigma_I=0.1m_{\rm pl}$ for two cases of 
$\phi_I=3m_{\rm pl}$ (solid) and $\phi_I=4m_{\rm pl}$
(dotted) with $d=6$ and $b_*=1.0 \times 10^5/m_{\rm pl}$.
When $\phi_I=3m_{\rm pl}$, inflation proceeds in the usual
manner while inflaton slowly evolves toward $\phi=0$.
When $\phi_I=4m_{\rm pl}$, the system evolves toward 
larger values of $\sigma$, and the value of inflaton hardly 
changes. {\bf Inset:} Evolution of the number of 
e-foldings $N$ during inflation for $\phi_I=3m_{\rm pl}$.
We find that $N>60$ is achieved.
\end{figcaption}
\end{center}
\end{figure}

\section{Preheating with extra dimensions}   

After inflation, the system enters the preheating stage
during which fluctuations of scalar fields will grow
by parametric resonance.
We introduce another massless scalar field $\chi$ 
coupled to inflaton, and adopt the following modified 
potential instead of $(\ref{B13})$:
\begin{eqnarray}
V(\phi,\chi)=\frac12 m^2\phi^2+
\frac12 g^2\phi^2\chi^2+\tilde{g}^2\phi^3\chi.
\label{C1}
\end{eqnarray}
Note that we include the interaction term 
$\tilde{g}^2\phi^3\chi$ which often appears
in supergravity models\cite{SUGRA} in addition to the 
standard term $\frac12 g^2\phi^2\chi^2$.
This provides a way to escape from an inflationary 
suppression of the $\chi$ field as we will show later. 

When we consider fluctuations of scalar fields, 
metric perturbations should be also taken into account 
for a consistent study of 
preheating\cite{mpre1,mpre2}.
In fact, inclusion of metric perturbations can change 
the evolution of field fluctuations significantly in broad
classes of models\cite{mpre7,mpre3,mpre4,mpre6}.
In this paper, we adopt the perturbed metric in the 
longitudinal gauge in the flat FRW background:
\begin{eqnarray}
ds^2=-(1+2\Phi)dt^2
+a^2(t)(1-2\Psi)\delta_{ij} dx^i dx^j,
\label{C2}
\end{eqnarray}
where $\Phi$ and $\Psi$ are gauge-invariant 
potentials\cite{MFB}. 

Decomposing scalar fields into 
homogeneous and gauge-invariant fluctuational
parts as $\phi(t,{\bf x}) \to \phi(t)+\delta\phi(t,{\bf x})$,
$\sigma(t,{\bf x}) \to \sigma(t)+\delta\sigma(t,{\bf x})$,
$\chi(t,{\bf x}) \to \chi(t)+\delta\chi(t,{\bf x})$,
and expanding scalar field fluctuations and metric 
fluctuations by Fourier modes, we obtain the
following perturbed equations (see e.g.,\cite{mpre2}):
\begin{eqnarray}
\Phi_k=\Psi_k,
\label{C3}
\end{eqnarray}
\begin{eqnarray}
\dot{\Phi}_k+H\Phi_k=\frac{\kappa^2}{2}
(\dot{\phi} \delta \phi_k+\dot{\sigma}
\delta\sigma_k+\dot{\chi} \delta\chi_k),
\label{C4}
\end{eqnarray}
\begin{eqnarray}
& &3H\dot{\Phi}_k + \left[\frac{k^2}{a^2}+
3H^2-\frac{\kappa^2}{2}(\dot\phi^2+
\dot\sigma^2+\dot\chi^2) \right] \Phi_k 
\nonumber \\
&=&
-\frac{\kappa^2}{2}\left(\dot{\phi}\delta\dot{\phi}_k
+U_{1,\phi}\delta\phi_k+\dot{\sigma}\delta\dot{\sigma}_k
+U_{1,\sigma}\delta\sigma_k+\dot{\chi}\delta\dot{\chi}_k
+U_{1,\chi}\delta\chi_k \right),
\label{C50}
\end{eqnarray}
\begin{eqnarray}
\delta\ddot{\phi}_k + 3H\delta\dot{\phi}_k+
\left[\frac{k^2}{a^2}+e^{-d\sigma/\sigma_*}
(m^2+g^2\chi^2+6\tilde{g}^2\phi\chi)\right]
\delta\phi_k 
= 4\dot{\phi} \dot{\Phi}_k+2(\ddot{\phi}
+3H\dot{\phi})\Phi_k-U_{1,\sigma\phi}
\delta\sigma_k-U_{1,\chi\phi}\delta\chi_k,
\label{C5}
\end{eqnarray}
\begin{eqnarray}
& &\delta\ddot{\sigma}_k + 3H\delta\dot{\sigma}_k+
\left[\frac{k^2}{a^2}+U''(\sigma)+
\frac{d^2}{\sigma_*^2}e^{-d\sigma/\sigma_*}
\left( \frac12 m^2\phi^2+\frac12 g^2\phi^2\chi^2
+\tilde{g}^2\phi^3\chi \right)
\right]\delta\sigma_k \nonumber \\
&=& 4\dot{\sigma}\dot{\Phi}_k+2(\ddot{\sigma}
+3H\dot{\sigma})\Phi_k-U_{1,\phi\sigma}
\delta\phi_k-U_{1,\chi\sigma}\delta\chi_k,
\label{C6}
\end{eqnarray}
\begin{eqnarray}
\delta\ddot{\chi}_k &+& 3H\delta\dot{\chi}_k+
\left( \frac{k^2}{a^2}+g^2\phi^2 e^{-d\sigma/\sigma_*}
\right) \delta\chi_k
= 4\dot{\chi}\dot{\Phi}_k+2(\ddot{\chi}
+3H\dot{\chi})\Phi_k-U_{1,\sigma\chi}
\delta\sigma_k-U_{1,\phi\chi}\delta\phi_k,
\label{C7}
\end{eqnarray}
where $U_{1,\phi\sigma}$, $U_{1,\chi\sigma}$, and 
$U_{1,\chi\phi}$ are expressed as
$U_{1,\phi\sigma}=-de^{-d\sigma/\sigma_*}
(m^2+g^2\chi^2+3\tilde{g}^2\phi\chi)\phi/\sigma_*$,
$U_{1,\chi\sigma}=
-de^{-d\sigma/\sigma_*}(g^2\chi+\tilde{g}^2
\phi)\phi^2/\sigma_*$, and
$U_{1,\chi\phi}=e^{-d\sigma/\sigma_*}(2g^2\phi\chi
+3\tilde{g}^2\phi^2)$, respectively.

The relation $(\ref{C3})$ indicates that the anisotropic 
stress vanishes at linear order.
Eliminating the $\dot{\Phi}_k$ term in Eqs.~$(\ref{C4})$
and $(\ref{C50})$, we find 
\begin{eqnarray}
\left(\frac{k^2}{a^2}-\frac{\kappa^2}{2}
\sum_{J}\dot{\varphi}_J^2 \right) \Phi_k =
-\frac{\kappa^2}{2}\sum_{J} \left(
\dot{\varphi}_J \delta\dot{\varphi}_{Jk}
+3H\dot{\varphi}_J \delta{\varphi}_{Jk}
+U_{1,\varphi_J}\delta \varphi_{Jk}\right),
\label{C51}
\end{eqnarray}
where $\varphi_J (J=1, 2, 3)$ correspond to the scalar 
fields $\phi$, $\sigma$, $\chi$, respectively.
Eq.~$(\ref{C51})$ shows that metric perturbations are 
known when evolutions of scalar fields are determined.
When field fluctuations are amplified, it is expected that 
this stimulates the growth of metric perturbations 
by Eq.~$(\ref{C51})$.
The enhancement of metric perturbations will also assist
the excitation of field perturbations as is found by  
Eqs.~$(\ref{C5})$-$(\ref{C7})$.

Parametric amplification of field fluctuations affects  
evolutions of the background quantities.
Since the $\chi$ fluctuation generally grows 
faster than other field fluctuations, we include this 
contribution in the background equations as
\begin{eqnarray}
\ddot{\phi}+3H \dot{\phi}+e^{-d\sigma/\sigma_*}
(m^2+g^2\langle\chi^2\rangle+
3\tilde{g}^2\phi\chi)\phi=0,
\label{C21}
\end{eqnarray}
\begin{eqnarray}
\ddot{\sigma}+3H \dot{\sigma}+U'(\sigma)
-\frac{d}{\sigma_*} e^{-d\sigma/\sigma_*}
\left(\frac12 m^2\phi^2+\frac12 g^2\phi^2
\langle\chi^2\rangle+\tilde{g}^2\phi^3\chi 
\right)=0,
\label{C22}
\end{eqnarray}
\begin{eqnarray}
\ddot{\chi}+3H \dot{\chi}+e^{-d\sigma/\sigma_*}
\left(g^2\phi^2\chi+\tilde{g}^2\phi^3 \right)=0,
\label{C23}
\end{eqnarray}
\begin{eqnarray}
H^2 =\frac{\kappa^2}{3}
\left[ \frac12 \dot{\sigma}^2+U(\sigma)
+\frac12 \dot{\phi}^2+\frac12 \dot{\chi}^2
+e^{-d\sigma/\sigma_*}
\left(\frac12 m^2\phi^2+\frac12 g^2\phi^2
\langle\chi^2\rangle+\tilde{g}^2
\phi^3\chi \right) \right],
\label{C24}
\end{eqnarray}
where the spatial average of the $\chi$ fluctuation is
defined by
\begin{eqnarray}
\langle\chi^2\rangle=
\frac{1}{2\pi^2} \int k^2 |\delta\chi_k|^2dk.
\label{C25}
\end{eqnarray}
Let us examine evolutions of the background 
fields and the scale factor.
As is found in the previous section, the $\sigma$ field
rapidly decreases during inflation compared with the 
$\phi$ field, and the condition $\sigma \ll \sigma_*=
\sqrt{d(d+2)/16\pi}~m_{\rm pl}$ holds 
at the beginning of preheating.
Then, in the stage where the $\chi$ fluctuation is not 
significantly enhanced, Eqs.~$(\ref{C21})$-$(\ref{C24})$
can be approximately written as
\begin{eqnarray}
\ddot{\phi}+3H \dot{\phi}+m^2\phi=0,
\label{C8}
\end{eqnarray}
\begin{eqnarray}
\ddot{\sigma}+3H \dot{\sigma}+
\frac{2(d-1)}{b_*^2}\sigma=0,
\label{C9}
\end{eqnarray}
\begin{eqnarray}
\ddot{\chi}+3H \dot{\chi}
+g^2\phi^2\chi+\tilde{g}^2\phi^3=0,
\label{C10}
\end{eqnarray}
\begin{eqnarray}
H^2=\frac{\kappa^2}{3}
 \left( \frac12 \dot{\phi}^2+
  \frac12 m^2\phi^2 \right).
\label{C11}
\end{eqnarray}
Making use of the time-averaged relation 
$\langle \frac12 \dot{\phi}^2\rangle_T=\langle \frac12
m^2\phi^2 \rangle_T$ during the oscillating stage of inflaton, 
the evolution of inflaton is analytically expressed by
Eqs.~$(\ref{C8})$ and $(\ref{C11})$ as 
\begin{eqnarray}
\phi=\Phi(t) \sin mt,~~~~{\rm with}~~~~ 
\Phi(t)=\frac{m_{\rm pl}}{\sqrt{3\pi}mt}.
\label{C41}
\end{eqnarray}
The coherent oscillation of inflaton begins when 
$\Phi (t_i) \sim 0.2m_{\rm pl}$, and we set the initial
time as $mt=\pi/2$ as in Ref.~\cite{KLS2}.
The scale factor evolves as $a \sim t^{2/3}$ since the 
system is dominated by the oscillation of the massive
inflaton field.
Although the $\sigma$ field oscillates with a frequency 
$\sqrt{2(d-1)}/b_*$, its amplitude is very small relative to
that of the $\phi$ field. For example, in the simulation
of Fig.~2, the amplitude of $\sigma$ at the start 
of preheating is found to be about $10^{-5}m_{\rm pl}$.

If we neglect metric perturbations, the evolution of the 
$\delta\chi_k$ fluctuation can be studied analytically 
at the linear stage of preheating.
Ignoring the r.h.s. of Eq.~$(\ref{C7})$ and introducing 
a new scalar field $\delta X_k=a^{3/2}\delta \chi_k$,
Eq.~$(\ref{C7})$ reads
\begin{eqnarray}
\frac{d^2}{dt^2} \delta X_k+
\left[ \frac{k^2}{a^2}+g^2\phi^2
-\frac34 \left(\frac{2\ddot{a}}{a}+
\frac{\dot{a}^2}{a^2} \right) 
\right] \delta X_k=0.
\label{C12}
\end{eqnarray}
The last term in Eq.~$(\ref{C12})$ which corresponds to the 
pressure term can be neglected during the oscillating stage
of inflaton.
Then Eq.~$(\ref{C12})$ is reduced to the well-known
Mathieu equation\cite{Mathieu},
\begin{eqnarray}
\frac{d^2}{d z^2} \delta X_k+ 
\left(A_k -2q \cos 2z \right) \delta X_k=0,
\label{C13}
\end{eqnarray}
where $z=mt$ and
\begin{eqnarray}
A_k= 2q + \frac{k^2}{(ma)^2},~~~~
q=\frac{g^2\Phi^2(t)}{4m^2}.
\label{C15}
\end{eqnarray}
Then the value of $q$ at the beginning of preheating 
is estimated as
\begin{eqnarray}
q_i \approx 10^{10} \times g^2,
\label{C16}
\end{eqnarray}
where we used $\Phi (t_i) \approx 0.2m_{\rm pl}$
and $m \sim 10^{-6}m_{\rm pl}$.
The coupling $g~\lsim~10^{-5}$ yields 
 $q_i~\lsim~1$, which is generally called 
the narrow resonance.
In this case, parametric resonance is weak
in an expanding universe.
However, when $q_i \gg 1$, it was pointed out in 
Ref.~\cite{KLS1} that the $\chi$ particle production can be 
efficient in spite of the decrease of $q$ due to cosmic 
expansion, which was later confirmed
by numerical calculations in Ref.~\cite{KTmassivehartree}.
In this case, the $\delta\chi_k$ field initially lies 
in the broad resonance regime as long as $k$ is not so large 
relative to $ma$, and it jumps over many stability 
and instability bands with the decrease of $q$.
This was termed {\it stochastic resonance} in Ref.~\cite{KLS2},
in which the $\delta\chi_k$ fluctuation increases 
stochastically overcoming the diluting effect by the expansion
of the universe.
For $g~\gsim~3 \times 10^{-4}$, the backreaction effect
of created $\chi$ particles becomes important,
which results in the termination of parametric resonance. 
In this case, since the coherent oscillation of the inflaton field
is broken  by the growth of the $\delta\chi_k$
fluctuation, the analytical
method based on the Mathieu equation is no longer applied. 
In this respect, several numerical works have been done
by making use of mean field approximations
or fully nonlinear calculations.
In the Hartree approximation the final variance of the 
$\chi$ field  is estimated as $\langle\chi^2\rangle_f
\propto q^{-1/2}$\cite{KTmassivehartree}, while in the fully 
nonlinear calculations it was found to be $\langle\chi^2\rangle_f
\propto q^{-1}$ for the case of $q \gg 1$\cite{KTmassivefull}.

As for the $\delta\sigma_k$ field, there exist resonance
terms in the l.h.s. of Eq.~$(\ref{C6})$ which may
lead to the enhancement of the fluctuation 
of dilaton. If we neglect the effect of metric perturbations 
in the r.h.s. of Eq.~$(\ref{C6})$ and 
making use of the relation $|\sigma| \ll \sigma_*$
during preheating, the equation of the 
$\delta\sigma_k$ field can be approximately written as
\begin{eqnarray}
\delta\ddot{\sigma}_k + 3H\delta\dot{\sigma}_k+
\left[\frac{k^2}{a^2}+\frac{2(d-1)}{b_*^2}
+\frac{8\pi d}{d+2}\left(\frac{m}
{m_{\rm pl}}\right)^2 \phi^2 \right]\delta\sigma_k=0.
\label{C17}
\end{eqnarray}
Defining a new scalar field $\delta \Sigma_k=
a^{3/2}\delta \sigma_k$ and ignoring the contribution 
from the pressure term, we obtain 
\begin{eqnarray}
\frac{d^2}{d z^2} \delta \Sigma_k+ 
\left(A_k -2q \cos 2z \right) \delta \Sigma_k=0,
\label{C18}
\end{eqnarray}
where 
\begin{eqnarray}
A_k= 2q + \frac{2(d-1)}{b_*^2m^2}
+\frac{k^2}{(ma)^2},~~~~
q=\frac{2d}{3(d+2)z^2}.
\label{C19}
\end{eqnarray}
For $d \ge 1$, $q_i~\lsim~1$ at the beginning of preheating.
Moreover, since $q$ decreases as $q \sim t^{-2}$,
parametric resonance is very weak.
Namely, in the unperturbed metric, 
analytic estimates indicate that 
the dilaton fluctuation does not grow during preheating.
We also find from Eq.~$(\ref{C5})$  that the 
enhancement of the inflaton fluctuation can not be
expected in the absence of metric perturbations.

Let us proceed to the case where metric perturbations
are taken into account. 
{}From Eq.~$(\ref{C4})$, we can expect that 
the growth of the $\delta\chi_k$ field will enhance 
metric perturbations.
On the other hand, it was pointed out in 
Refs.~\cite{mpre8,mpre10,mpre11,mpre12}
that the amplitude of super-Hubble fluctuations in the 
$\delta\chi_k$ field is severely damped during inflation
in the case where $g\phi$ is much larger than the Hubble
expansion rate $H$ with a model of 
$V(\phi,\chi)=\frac12 m^2\phi^2+\frac12 g^2\phi^2\chi^2$.
Later, Bassett {\it et al.}\cite{mpre3}
showed that inclusion of the interaction
$\tilde{g}^2\phi^3\chi$ protects super-Hubble 
$\delta\chi_k$ fluctuations from being suppressed.
In what follows, we will consider both cases of $\tilde{g}=0$
and $\tilde{g} \ne 0$ separately.

\subsection{Case of $\tilde{g}=0$}   
Let us first estimate the amplitude of super-Hubble 
$\delta\chi_k$ modes at the beginning of preheating.
When $\tilde{g}=0$, the adiabatic solution  for
$\delta\chi_k$ during inflation is expressed as
\begin{eqnarray}
\delta\chi_k=\frac{a^{-3/2}}{\sqrt{2\omega_k}}
e^{-i\omega_k t},
\label{C60}
\end{eqnarray}
where $\omega_k^2=k^2/a^2+g^2\phi^2 e^{-d \sigma/\sigma_*}
\approx k^2/a^2+g^2\phi^2$.
In order to lead to efficient $\chi$ particle production,
the resonance parameter is required to be 
$q=g^2\phi^2/4m^2 \gg 1$.
In this case, the effective mass of 
the $\delta\chi_k$ field is much larger than the Hubble
expansion rate $H \sim m$ during inflation.
Then the amplitude of the super-Hubble 
$\delta\chi_k$ field for modes 
relevant for structure formation is estimated 
as $|\delta\chi_k| \sim a^{-3/2}/\sqrt{g\phi}$, 
which exponentially decreases during inflation.
On the other hand, since the effective mass of 
the $\delta\phi_k$ field in the l.h.s. of Eq.~$(\ref{C5})$
is comparable to the Hubble rate $H$, the super-Hubble
inflaton fluctuation is not affected 
by the suppression during inflation.

As for the $\delta\sigma_k$ field, its effective mass 
for super-Hubble modes is given by Eq~$(\ref{C17})$ as
\begin{eqnarray}
m_{\rm eff}^2=\left[ \frac{2(d-1)}{b_*^2m^2}+
\frac{8\pi d}{d+2}\left(\frac{\phi}
{m_{\rm pl}}\right)^2 \right]m^2.
\label{C61}
\end{eqnarray}
For the typical scale $b_*$ which is determined by 
Eq.~$(\ref{B16})$ and the initial value of inflaton 
$\phi_I~\gsim~3m_{\rm pl}$, $m_{\rm eff}$ 
is estimated as $m_{\rm eff}^2~\gsim~100m^2\sim 10H^2$.
Hence the super-Hubble $\delta\sigma_k$ fluctuation
will be also affected by the inflationary suppression,
which is relevant for small $b_*$
and large initial values of $\phi$.
We can roughly estimate the amplitude of super-Hubble
$\delta\sigma_k$ modes by Eq.~$(\ref{C6})$ during inflation.
Neglecting the contributions of the $\chi$ field and the time
derivative terms of $\sigma$ and $\delta\sigma_k$, we obtain
the amplitude of $\delta\sigma_k$ at the start of preheating as
\begin{eqnarray}
|\delta \sigma_k(t_i)| \approx \frac{2}{(d-1)}
\sqrt{\frac{\pi d}{d+2}} \frac{\phi}{m_{\rm pl}}(b_* m)^2 
|\delta \phi_k(t_i)|.
\label{C62}
\end{eqnarray}
For example, when $d=6$, since $b_*$ is constrained as
$b_*~\lsim~1.1 \times 10^{5}/m_{\rm pl}$, 
 Eq.~$(\ref{C62})$ yields
\begin{eqnarray}
|\delta \sigma_k(t_i)| ~\lsim~10^{-3} |\delta \phi_k(t_i)|.
\label{C63}
\end{eqnarray}
This means that the suppression effect of super-Hubble modes
is weak compared with the $\delta\chi_k$ field
as long as $b_*$ is not much smaller than its upper bound. 

Let us estimate the impact on metric perturbations by the 
growth of field fluctuations.
First, we introduce the power spectrum of $\Phi_k$:
\begin{eqnarray}
{\cal P}(k)=\frac{k^3}{2\pi^2}|\Phi_k|^2=
\frac{|\tilde{\Phi}_k|^2}{2\pi^2},
\label{C42}
\end{eqnarray}
where $\tilde{\Phi}_k \equiv k^{3/2} \Phi_k$.
Defining new scalar fields 
$\tilde{\varphi}_{J} \equiv \varphi_{J}/m_{\rm pl}$
and $\delta \tilde{\varphi}_{Jk} \equiv k^{3/2}\delta 
\varphi_{Jk}/m_{\rm pl}$ ($J=1, 2, 3$), 
we obtain the following relation from 
Eq.~$(\ref{C4})$:
\begin{eqnarray}
\dot{\tilde{\Phi}}_k+H\tilde{\Phi}_k=
4\pi(\dot{\tilde{\phi}} \delta\tilde{\phi}_k+
\dot{\tilde{\sigma}}
\delta\tilde{\sigma}_k+\dot{\tilde{\chi}}
\delta\tilde{\chi}_k).
\label{C20}
\end{eqnarray}
For super-Hubble modes $k \ll aH$,
the amplitude of $\delta\tilde{\chi}_k$ is written as
$|\delta\tilde{\chi}_k| \approx \bar{k}^{3/2}(m/
m_{\rm pl}) \sqrt{m/(2g\phi)}$ where $\bar{k}
\equiv k/(ma_i)$ with $a_i$ the scale factor
at the onset of preheating.
Since the cosmological modes correspond to 
$\bar{k}\sim e^{-60} \sim 10^{-26}$, 
$|\delta\tilde{\chi}_k|$ is estimated as 
$|\delta\tilde{\chi}_k|~\lsim~10^{-45}$ for 
the broad resonance case $q\gg 1$. 
The homogeneous part of the $\chi$ field 
is also affected by this strong suppression
[see Eq.~$(\ref{C23})$ with $\tilde{g}=0$]. 
At the beginning of preheating, the $\dot{\chi}\delta\chi_k$ 
term in the r.h.s. of Eq.~$(\ref{C4})$ is very small 
relative to the $\dot{\phi}\delta\phi_k$ term.
Although super-Hubble $\delta\chi_k$ fluctuations
exhibit parametric amplification during preheating, 
it increases only by the factors $10^4-10^5$
for the coupling of $g=3\times 10^{-4}-10^{-2}$\cite{mpre10}.
Hence we can expect that the excitement of the $\delta\chi_k$
fluctuation hardly affects the evolution 
of super-Hubble metric perturbations by analytic estimates.

\begin{figure}
\begin{center}
\singlefig{12cm}{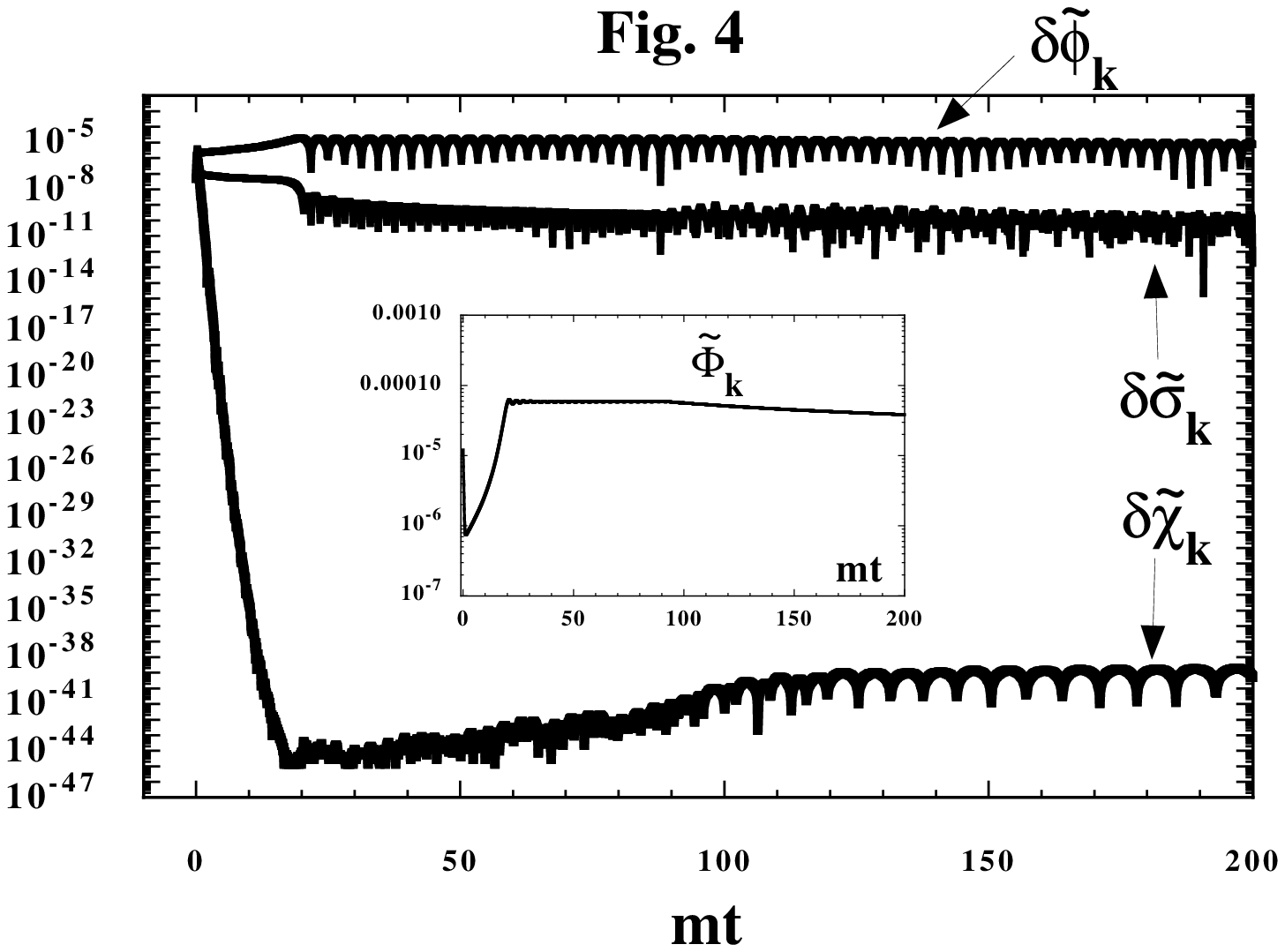}
\begin{figcaption}{Fig4}{12cm}
The evolutions of field perturbations
$\delta\tilde{\phi}_k$, $\delta\tilde{\sigma}_k$, 
$\delta\tilde{\chi}_k$
for a super-Hubble mode $\bar{k}=10^{-26}$ with 
$g=1.0 \times 10^{-3}$, $d=6$,
$b_*=1.0 \times 10^5/m_{\rm pl}$, 
and $m=10^{-6}m_{\rm pl}$. 
See the text for the initial 
conditions of scalar fields. Note that we start integrating 
about 60 e-foldings before the beginning of preheating.
{\bf Inset:} The evolution of the metric 
perturbation $\tilde{\Phi}_k$ for a super-Hubble 
mode $\bar{k}=10^{-26}$. 
\end{figcaption}
\end{center}
\end{figure}

We are also concerned with whether super-Hubble 
$\delta\sigma_k$ fluctuations are enhanced or not
during preheating.
Although the inflationary suppression for $\delta\sigma_k$ 
is not so significant as compared with the $\delta\chi_k$ case,
we have to keep in mind that $\delta\sigma_k$
can not be enhanced unless metric perturbations are 
taken into account.
In order to amplify super-Hubble metric perturbations,
we generally require some scalar fields such as $\chi$
which exhibit parametric amplification even in the absence of 
metric perturbations.
However, when $\tilde{g}=0$, the 
$\delta\chi_k$ fluctuation in large scales is strongly suppressed.
It is expected that the super-Hubble fluctuation of dilaton 
will be held static during preheating, because  
the $\sigma$ field will not play a dominant
role to stimulate the enhancement of metric perturbations.

In order to verify the above estimates, we numerically solved 
perturbed equations
$(\ref{C3})$-$(\ref{C7})$ along with background 
equations $(\ref{C21})$-$(\ref{C24})$. 
In Fig.~4, we plot the evolutions of field perturbations
$\delta\tilde{\phi}_k$, $\delta\tilde{\sigma}_k$,
$\delta\tilde{\chi}_k$, 
and the metric perturbation $\tilde{\Phi}_k$ during inflation and
preheating for a super-Hubble mode 
$\bar{k}=10^{-26}$ with $g=1.0 \times 10^{-3}$, $d=6$,
$b_*=1.0 \times 10^5/m_{\rm pl}$, 
and $m=10^{-6}m_{\rm pl}$.
The initial values of homogeneous scalar fields
are chosen as $\phi_I=3.0m_{\rm pl}$, 
$\sigma_I=0.1m_{\rm pl}$, and $\chi_I=1.0 \times
10^{-3} m_{\rm pl}$. As for the initial field perturbations,
we take
\begin{eqnarray}
|\delta\varphi_{Jk}| =\frac{1}{\sqrt{2\omega_{Jk}}},~~~
|\delta\dot{\varphi}_{Jk}| =\omega_{Jk}|\delta\varphi_{Jk}|,
\label{C64}
\end{eqnarray}
where $\omega_{Jk}^2\equiv k^2/a^2+m_{\varphi_J}^2$
($J=1,2,3)$ with $m_{\varphi_J}$ is the effective mass
of the each scalar field in the l.h.s. of 
Eqs.~$(\ref{C5})$-$(\ref{C7})$.

In Fig.~4, the $\delta\chi_k$ fluctuation  is exponentially 
suppressed during inflation
($0<mt~\lsim~20$), yielding $\delta\tilde{\chi}_k \sim 10^{-45}$
at the beginning of preheating.
Although $\delta\chi_k$ is enhanced 
by parametric resonance for $mt~\gsim~20$, the final value is
very small as $\delta\tilde{\chi}_k \sim 10^{-40}$.
The amplitude of $\delta\sigma_k$ is by three 
orders of magnitude smaller than that of $\delta\phi_k$
as is analytically estimated by Eq.~$(\ref{C62})$
at the end of inflation. We also find in Fig.~4 
that both of super-Hubble
$\delta\sigma_k$ and $\delta\phi_k$ fluctuations do not 
grow during preheating, which means that field fluctuations
are not assisted by the presence of metric perturbations.
As is found in the inset of Fig.~4, super-Hubble 
metric perturbations remain almost constant 
during preheating.
In the case of $\tilde{g}=0$, 
the usual prediction of the inflationary 
spectrum in large scales is not likely to be modified, and 
the analysis neglecting metric perturbations gives almost 
the same results as compared with the perturbed metric case.
As a result, the fluctuation of dilaton on super-Hubble
scales can not be amplified during preheating.

Let us next consider smaller scales which are within the 
Hubble radius at the beginning of preheating.
Since this corresponds to the modes $k~\gsim~a_iH_i$,
the condition $k^2/a^2>g^2\phi^2$ holds in most stage
of inflation. Hence the amplitude of these
modes during inflation is approximately expressed 
by Eq.~$(\ref{C60})$ as
\begin{eqnarray}
|\delta \chi_k| \approx \frac{1}{a\sqrt{2k}}.
\label{C65}
\end{eqnarray}
The r.h.s. of Eq.~$(\ref{C65})$ decreases slower compared 
with the super-Hubble modes. In fact, for the modes of
$k^2/a_i^2~\gsim~g^2\phi_i^2$ at the start of preheating,
we obtain
\begin{eqnarray}
|\delta \tilde{\chi}_k (t_i)| \approx \frac{1}{\sqrt{2}}
\frac{m}{m_{\rm pl}} \bar{k}.
\label{C66}
\end{eqnarray}
For sub-Hubble modes $\bar{k}=k/(ma_i)~\gsim~1$, we find 
$|\delta \tilde{\chi}_k (t_i)|~\gsim~10^{-6}$, which is much 
larger than in the super-Hubble case.
However, since the homogeneous part of the $\chi$ field is severely
damped, the $\dot{\chi}\delta\chi_k$ term in the r.h.s. of 
Eq.~$(\ref{C4})$ is still much smaller than the  
$\dot{\phi}\delta\phi_k$ term at the beginning of preheating.
This indicates that parametric amplification of 
the $\delta\chi_k$ fluctuation will not lead to the 
growth of sub-Hubble metric perturbations.
We have numerically confirmed that metric perturbations and 
the $\delta\sigma_k$ fluctuation for sub-Hubble modes
$1~\lsim~k/(ma_i)~\lsim~100$ are not relevantly enhanced 
for the coupling of $3 \times 10^{-4}~\lsim~g~\lsim~10^{-2}$
(see Fig.~5).
However, we have to caution that including the second order
metric backreaction effect\cite{SMP} will lift the homogeneous $\chi$ field,
which may assist the enhancement of metric and field fluctuations 
in sub-Hubble modes as was mentioned in Ref.~\cite{mpre10}.
The full backreaction issues are left for the future work.

\begin{figure}
\begin{center}
\singlefig{12cm}{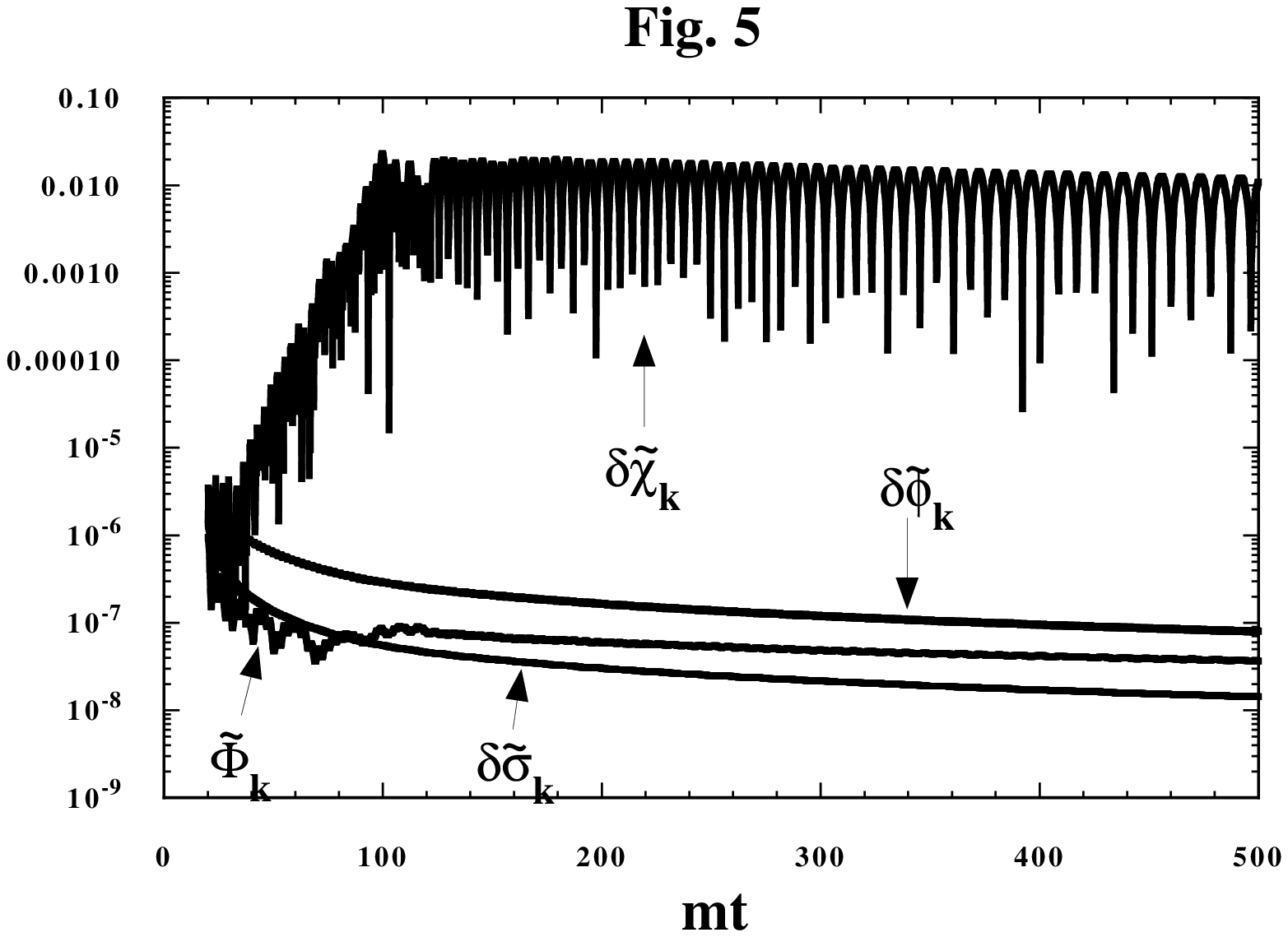}
\begin{figcaption}{Fig5}{12cm}
The evolutions of field perturbations
$\delta\tilde{\phi}_k$, $\delta\tilde{\sigma}_k$,
$\delta\tilde{\chi}_k$, and the metric perturbation
$\tilde{\Phi}_k$ for a sub-Hubble mode $\bar{k}=5$ with 
$g=1.0 \times 10^{-3}$, $d=6$,
$b_*=1.0 \times 10^5/m_{\rm pl}$, 
and $m=10^{-6}m_{\rm pl}$. 
Note that we start integrating from 
the beginning of preheating.
\end{figcaption}
\end{center}
\end{figure}

In the case of $\tilde{g}=0$, we argue that 
the dilaton fluctuation as well as metric perturbations
in both super- and sub-Hublle scales 
can not be strongly amplified during preheating.

\subsection{Case of $\tilde{g} \ne 0$}   
 
If the coupling $\tilde{g}^2\phi^3\chi$ is taken into account,
the suppression of the super-Hubble $\delta\chi_k$ fluctuation 
can be avoided. For super-Hubble modes, neglecting 
derivative terms of $\delta\chi_k$ and $\chi$
fields as well as the $\delta\sigma_k$ term
in Eq.~$(\ref{C7})$, we find the following relation 
at the end of inflation:
\begin{eqnarray}
\delta \chi_k \approx -3\left(\frac{\tilde{g}}{g}\right)^2
\delta \phi_k.
\label{C67}
\end{eqnarray}
As for the homogeneous part of the $\chi$ field,
Eq.~$(\ref{C10})$ implies
\begin{eqnarray}
\chi \approx -\left(\frac{\tilde{g}}{g}\right)^2 \phi.
\label{C68}
\end{eqnarray}
We find from Eqs.~$(\ref{C67})$ and $(\ref{C68})$
that both of super-Hubble $\delta\chi_k$ fluctuations and the 
homogeneous $\chi$ field are not severely
suppressed compared with the $\tilde{g}=0$ case.
Then we can expect that the growth of the $\dot{\chi}
\delta{\chi}_k$ term in the r.h.s. of Eq.~$(\ref{C4})$
during preheating may lead to the enhancement
of super-Hubble metric perturbations. 

Although larger values of $\tilde{g}$ will surely 
escape the inflationary suppression, we have to take
care that this may prevent the successful 
inflationary scenario. 
During inflation, the frequency $\Omega_{\phi}$
of the inflaton condensate is estimated by making use of 
Eq.~$(\ref{C68})$ as 
\begin{eqnarray}
\Omega_{\phi}^2 \equiv 
m^2+g^2\chi^2+3e^{-d\sigma/\sigma_*}
\tilde{g}^2\phi\chi \approx m^2\left[ 
1-\frac{2\tilde{g}^2\phi^2}
{m^2} \left(\frac{\tilde{g}}{g}\right)^2 \right],
\label{C69}
\end{eqnarray}
where we used the relation $e^{-d\sigma/\sigma_*}
\approx 1$.
Eq.~$(\ref{C69})$ indicates that larger values of $\tilde{g}$
make $\Omega_{\phi}^2$ negative, and will 
lead to the unphysical result that $\phi$ increases by 
negative instability.
In order to avoid this, 
$\Omega_{\phi}^2$ should be positive 
at the start of inflation ($\phi_I \sim 3m_{\rm pl}$). 
Then the ratio $\tilde{g}/g$  is constrained as 
\begin{eqnarray}
\frac{\tilde{g}}{g}~\lsim~\frac{5\times 10^{-4}}
{\sqrt{g}},
\label{C70}
\end{eqnarray}
where we used $m \sim 10^{-6}m_{\rm pl}$.
For an efficient $\chi$ particle production, the 
coupling $g$ is required to be $g~\gsim~3.0 \times10^{-4}$
\cite{KLS2}, which leads to the constraint:
$\tilde{g}/g~\lsim~2.9 \times 10^{-2}$.
The upper bound of $\tilde{g}/g$ decreases 
with the increase of $g$ as is found by Eq.~$(\ref{C70})$.
For example, $\tilde{g}/g~\lsim~1.6 \times 10^{-2}$
for $g=1.0 \times 10^{-3}$, and 
$\tilde{g}/g~\lsim~5.0 \times 10^{-3}$
for $g=1.0 \times 10^{-2}$.
Although larger values of $g$ are favorable for the 
rapid growth of the $\delta\chi_k$ fluctuation, 
this simultaneously results in the stronger suppression
for $\delta\chi_k$ and $\chi$ in Eqs.~$(\ref{C67})$
and $(\ref{C68})$.

We have numerically examined the dynamics of 
preheating in the coupling regimes 
$g=3 \times 10^{-4}-10^{-2}$ with $\tilde{g}$
constrained by Eq.~$(\ref{C70})$, and also
checked that the inflationary period proceeds in the 
usual manner.
In Fig.~6, the evolutions of field perturbations
$\delta\phi_k$, $\delta\chi_k$, $\delta\sigma_k$ and 
the metric perturbation $\Phi_k$ are
depicted for a super-Hubble mode 
$\bar{k}=10^{-26}$ with $g=5.0 \times 10^{-4}$,
$\tilde{g}/g=2.0 \times 10^{-2}$.
We start integrating about 60
e-foldings before the beginning of preheating, and choose
initial conditions $\phi_I=3.0m_{\rm pl}$,
$\sigma_I=0.1m_{\rm pl}$, and $\chi_I=-(\tilde{g}/g)^2\phi_I$
for the homogeneous part, and use Eq.~$(\ref{C64})$ for
the fluctuational parts.
In this case, the analytic estimation $(\ref{C67})$ implies
the relation $|\delta \chi_k(t_i)| \approx 10^{-3}|\delta \phi_k(t_i)|$ 
at the onset of preheating ($mt \approx 20$),
which can be easily confirmed in Fig.~6(a).
The super-Hubble $\delta\chi_k$ fluctuation starts to grow
from $mt \approx 30$ by parametric resonance, and 
catches up the $\delta\phi_k$ fluctuation at $mt \approx 120$.
At this stage, the backreaction effect of the produced
$\chi$ particle begins to destroy the coherent oscillation 
of the $\phi$ field [see the inset of Fig.~6(a)].
In spite of this, the amplification of the $\delta\chi_k$ fluctuation 
still takes place before the oscillation 
of $\phi$ is completely broken at $mt \approx 220$.
For  $140~\lsim~mt~\lsim~170$, 
the super-Hubble $\delta\phi_k$ fluctuation is enhanced 
by about two orders of magnitude.
This occurs in the perturbed metric case where the r.h.s. of
Eq.~$(\ref{C5})$ stimulates the excitement of the 
$\delta\phi_k$ fluctuation.
However, since the increase of $\delta\chi_k$ 
is weakened by the backreaction effect of created particles,
the period during which the $\delta\phi_k$ fluctuation 
is enhanced does not continue long.
In Fig.~6(b), we find that the super-Hubble metric 
perturbation $\Phi_k$ begins to oscillate for $mt~\gsim~180$,
which is due to the enhancement of field fluctuations.
However, $\Phi_k$ does not increase even by one order 
of magnitude from the beginning of preheating.
This is mainly because the 
backreaction effect restricts the rapid increase 
of $\delta\chi_k$ soon after the super-Hubble 
$\delta\chi_k$ fluctuation catches up  
$\delta\phi_k$. Although one may think that
larger values of $\tilde{g}$ will lead to the strong 
amplification of $\Phi_k$,
Eq.~$(\ref{C70})$ constrains the coupling 
as $\tilde{g}/g~\lsim~2.2 \times 10^{-2}$ with 
$g=5.0\times 10^{-4}$, in which case 
the super-Hubble metric perturbation can not be strongly
excited. 

\begin{figure}
\begin{center}
\singlefig{12cm}{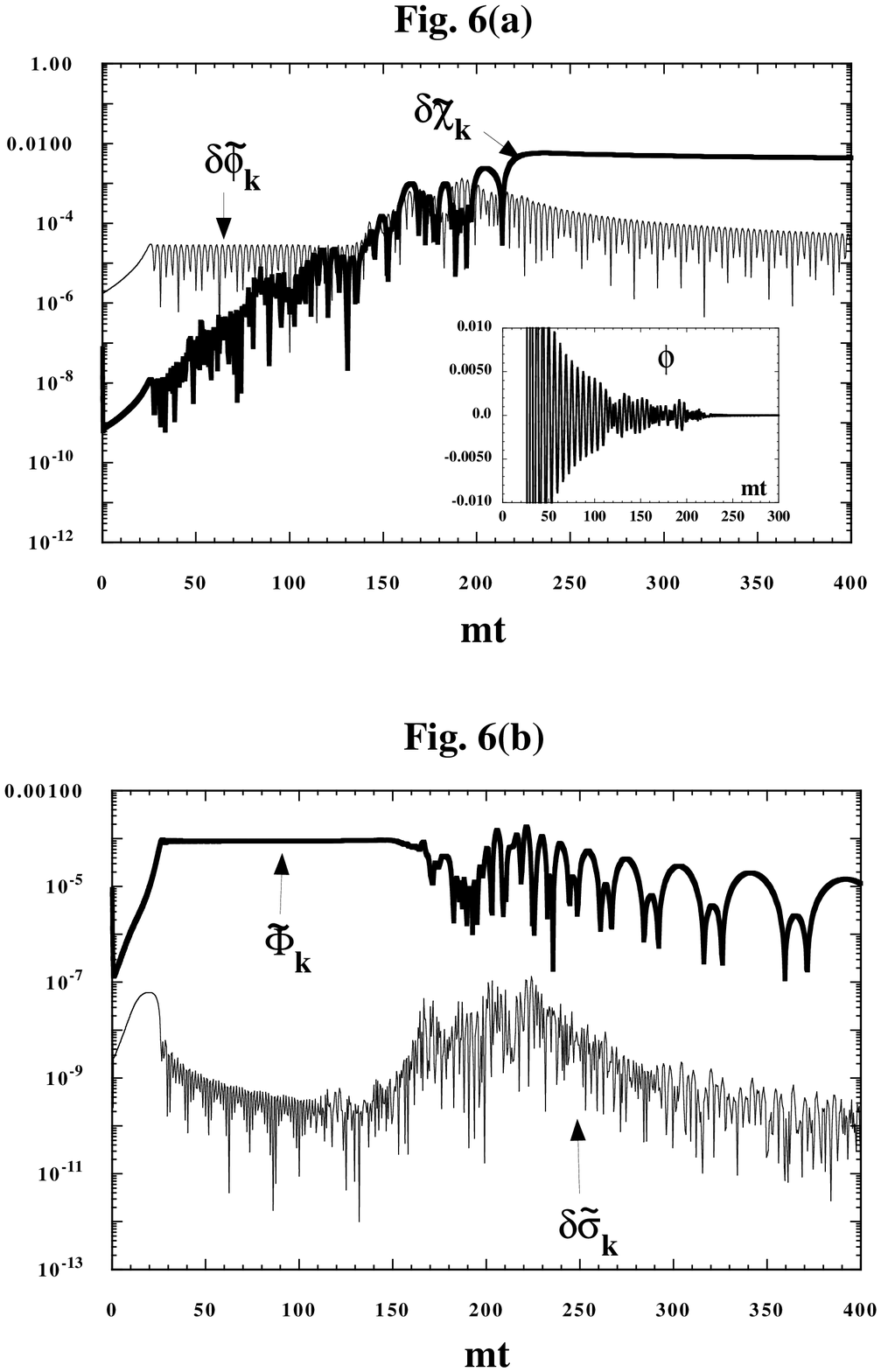}
\begin{figcaption}{Fig6}{12cm}
The evolutions of field perturbations
$\delta\tilde{\phi}_k$, $\delta\tilde{\chi}_k$
$\delta\tilde{\sigma}_k$, and the metric perturbation
$\tilde{\Phi}_k$ for a super-Hubble
mode $\bar{k}=10^{-26}$
with $g=5.0 \times 10^{-4}$, $\tilde{g}/g=2.0\times
10^{-2}$, $d=6$, $b_*=1.0 \times 10^5/m_{\rm pl}$, 
and $m=10^{-6}m_{\rm pl}$ 
during inflation and preheating. 
{\bf Inset of Fig.~6(a):} The evolution of the inflaton 
 condensate $\phi$.
\end{figcaption}
\end{center}
\end{figure}

We have also investigated other values 
of the coupling $g$, and numerical results exhibit the similar behavior.
The large scale cosmic microwave background (CMB) 
anisotropies will not be significantly modified with the existence
of the preheating phase,
even taking into account the coupling $\tilde{g}^2\phi^3\chi$.
However, the additional enhancement of the 
super-Hubble metric perturbation found in
Fig.~6(b) for $mt~\gsim~180$ may give some small imprints 
in the CMB spectrum.

The $\delta\sigma_k$ fluctuation in super-Hubble scales 
can be amplified a little in a short stage 
as in the case of $\delta\phi_k$.
In Fig.~6(b), we find that $\delta\sigma_k$ increases 
by about  two orders of magnitude during 
$140~\lsim~mt~\lsim~220$. 
However, for $g~\gsim~3.0 \times 10^{-4}$
and values of $\tilde{g}$ which satisfy the condition of 
$(\ref{C70})$, numerical calculations imply that 
the growth of super-Hubble $\delta\sigma_k$ 
modes is hardly expected except in the case where $\tilde{g}/g$
is close to its upper bound.
Even when $\tilde{g}/g$ is close to its upper bound as in the 
case of $g=5.0 \times 10^{-4}$, $\tilde{g}/g=2.0 \times 10^{-2}$,
the enhancement is found to be weak. Moreover, 
final fluctuations $\delta\tilde{\sigma}_k$ in super-Hubble 
modes are typically smaller than $\delta\tilde{\phi}_k$
and $\delta\tilde{\chi}_k$ as is found in Fig.~6.

\begin{figure}
\begin{center}
\singlefig{12cm}{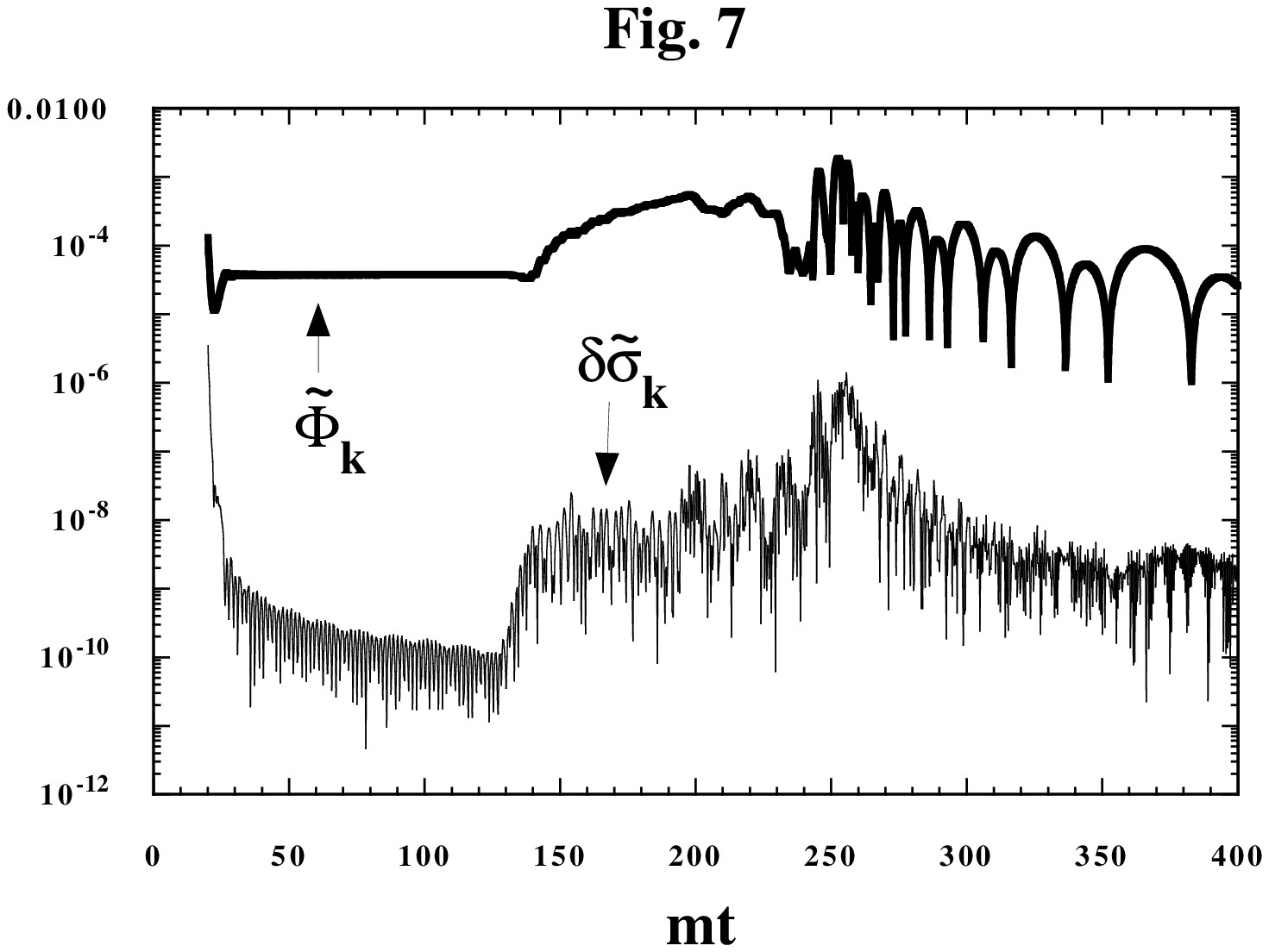}
\begin{figcaption}{Fig7}{12cm}
The evolutions of the metric perturbation 
$\Phi_k$ and the field perturbation $\delta\sigma_k$ for a 
sub-Hubble mode $\bar{k}=5$ with 
$g=5.0 \times 10^{-4}$, $\tilde{g}/g=2.0\times
10^{-2}$, $d=6$, $b_*=1.0 \times 10^5/m_{\rm pl}$, 
and $m=10^{-6}m_{\rm pl}$. 
\end{figcaption}
\end{center}
\end{figure}

As for sub-Hubble scales, amplifications of metric 
perturbations and the $\delta\sigma_k$ fluctuation are relevant
compared with the super-Hubble case.
In Fig.~7, we show the evolutions of $\Phi_k$ 
and $\delta\sigma_k$ during preheating for
a sub-Hubble mode $\bar{k}=5$
with $g=5.0 \times 10^{-4}$, $\tilde{g}/g=2.0\times 10^{-2}$.
In this case, the sub-Hubble fluctuation of $\delta\tilde{\chi}_k$ is 
larger than in the super-Hubble case at the start of preheating.
We find in Fig.~7 that $\Phi_k$ 
increases by more than one order of magnitude, which 
indicates that metric preheating can be vital in small scales 
in the presence of the $\tilde{g}^2\phi^3\chi$ term.
The dilaton fluctuation in sub-Hubble modes is also enhanced
with the growth of metric perturbations. However,
the final $\delta\sigma_k$ fluctuation 
does not exceed its fluctuation at the onset of preheating.
If we choose smaller values of $\tilde{g}$, the enhancement
of sub-Hubble $\Phi_k$ and $\delta\sigma_k$ modes becomes 
weaker. 
For the couplings which range 
$3 \times 10^{-4}~\lsim~g~\lsim~ 10^{-2}$,
we numerically find that 
the sub-Hubble dilaton fluctuation is not relevantly amplified 
for the value of $\tilde{g}$ which is smaller 
by one order of magnitude than its upper bound given 
by Eq.~$(\ref{C70})$.
When the standard coupling $g^2\phi^2\chi^2$
dominates over the coupling $\tilde{g}^3\phi^3\chi$ (namely
$g \gg \tilde{g}$), we conclude that the fluctuation of dilaton 
can be held static both in the sub- and super-Hubble scales.

\section{Concluding remarks and discussions}   
In this paper, we have studied preheating after inflation 
with a quadratic inflaton potential $V(\phi)=\frac12 m^2\phi^2$
in the presence of a dilaton field $\sigma$ which represents
the scale of compactifications in a higher-dimensional 
generalized Kaluza-Klein theory.
We consider the Candelas-Weinberg model
where extra dimensions are compactified on the sphere
with a cosmological constant and a one-loop quantum
correction (Casimir effects).
In the chaotic inflation model, 
a potential barrier which prevents the growth 
of the internal space disappears for large values 
of inflaton. However, the fine-tuned initial conditions
and parameters of the model naturally lead to successful inflation. 
We find that the existence of dilaton during inflation hardly
affects the evolution of inflaton, and the chaotic inflationary 
scenario proceeds in the usual manner as long as initial 
conditions are chosen so that 
dilaton does not go beyond the potential barrier.

At the stage of preheating after inflation,
another scalar field $\chi$ coupled to inflaton 
can be amplified by parametric resonance 
due to the oscillation of inflaton. In addition to the standard 
coupling $\frac12 g^2\phi^2\chi^2$, we have also included
the coupling $\tilde{g}^2\phi^3\chi$ by which the exponential 
suppression of the super-Hubble $\chi$ fluctuation can be avoided
during inflation. We include metric perturbations explicitly
for scalar field equations, and investigate how the fluctuation of 
dilaton will be amplified both in super- and sub-Hubble scales. 
Neglecting metric perturbations, the equation for the dilaton
fluctuation is reduced to the form of Mathieu equation 
as in the case of the $\chi$ fluctuation at the linear stage of preheating.
Since the resonance parameter $q$ is smaller
than unity during the whole stage of preheating, 
the dilaton fluctuation does not exhibit parametric amplification 
in the rigid spacetime case.

In the perturbed metric case, it is generally expected that field resonances
will stimulate the enhancement of metric perturbations $\Phi_k$.
In the case of $\tilde{g}=0$, however, since
low momentum modes of the $\chi$ field fluctuation are
severely suppressed during inflation, 
super-Hubble metric perturbations are hardly affected 
by parametric amplification of the field perturbation.
We have numerically verified that super-Hubble 
metric perturbations remain almost constant during preheating,
and also found that the dilaton fluctuation in super-Hubble modes
can not be enhanced.
As for sub-Hubble modes with $\tilde{g}=0$, the $\chi$ field 
fluctuation is not suppressed relative to super-Hubble modes.
However, since the source term in the equation of $\Phi_k$
contains the time derivative of the homogeneous $\chi$ field,
we can not expect the strong amplification of 
dilaton fluctuation as well as metric perturbations.

If the coupling $\tilde{g}^2\phi^3\chi$ is taken into account,
both the homogeneous and super-Hubble fluctuational parts
of the $\chi$ field can escape the inflationary suppression.
These are about $(\tilde{g}/g)^2$ times  those of inflaton
at the onset of preheating.
In this case, the $\chi$ fluctuation can typically increase to 
the order of the inflaton fluctuation, after which 
the backreaction effect of created $\chi$
particles onto the $\phi$ field begins to be relevant.
This restricts the further excitement of the $\chi$ fluctuation, and 
amplifications of the super-Hubble metric perturbation and 
the dilaton fluctuation are found to be weak even for large values of 
$\tilde{g}$ which are close to its upper bound.
As for sub-Hubble scales, the dilaton fluctuation can be modestly 
amplified by the growth of metric perturbations.
However, the dilaton fluctuation does not grow for the value
of $\tilde{g}$ which is smaller by one order of magnitude than its
upper bound given by Eq.~$(\ref{C70})$.
We argue that the stability of compactifications can be 
preserved during preheating in the quadratic chaotic inflationary
scenario as long as $\tilde{g} \ll g$.

We have found that the enhancement of the dilaton perturbation
is intimately related with the generation of metric perturbations.
Although we only considered the backreaction due to 
 field fluctuations, we should also include second order 
metric perturbations to the background equations 
for a consistent study. In fact, the effective momentum tensor
formalism in Ref.~\cite{SMP} gives rise to the coupling of
the metric $\Phi$ and the fluctuation $\delta\chi_k$
in the equation of the homogeneous $\phi$ field,
which leads to the growth of the super-Hubble $\delta\chi_k$
fluctuation as well as the homogeneous 
$\chi$ field. Although it is expected that  
metric perturbations in super-Hubble scales remain well within
linear regimes in the case of $\tilde{g}=0$ as 
in Ref.~\cite{mpre10}, inclusion of 
second order metric backreaction effects may lead to the distortion
of the large-scale CMB spectrum as well as the enhancement 
of sub-Hubble metric and field fluctuations in the presence of the interaction 
$\tilde{g}^2\phi^3\chi$. 
In addition to this, rescattering effects of field and metric 
fluctuations (i.e. mode-mode coupling) will be important at the 
final stage of preheating\cite{KLS2,KTselffull,KTmassivefull,mpre13}.
Although we do not consider these complicated issues 
in this paper, we should take into account the full backreaction
and the rescattering effects for a complete study of preheating.
 
In this paper, we have restricted ourselves to the massive 
inflaton model as a first step toward 
understanding  the dynamics of preheating with extra dimensions.
We found that amplification of the super-Hubble dilaton
fluctuation is weak in this model. 
However, in other models of inflation, there may
be a possibility that extra dimensions will be unstable during preheating.
Indeed, it was recently suggested that 
super-Hubble metric perturbations can be strongly enhanced 
in broad classes of models\cite{mpre7,mpre3,mpre6}.
One of such models is the massless chaotic inflation model
$V(\phi)=\frac14\lambda\phi^4$ with the interaction
$\frac12 g^2\phi^2\chi^2$. In this model, 
even when the effective mass of the $\chi$ field 
is comparable to the Hubble expansion rate $H$,
the super-Hubble
$\chi$ field fluctuation can be excited by 
parametric resonance.
This  will lead to the enhancement of the dilaton
fluctuation in long wave modes through
the amplification of super-Hubble metric perturbations.  
It was also pointed out in Ref.~\cite{mpre3} 
that the negative coupling $\frac12 g^2\phi^2\chi^2$ with 
$g^2<0$\cite{negative}
or the negative nonminimal coupling $\frac12 \xi R\chi^2$ with 
$\xi<0$\cite{nonminimalpre} will escape 
the inflationary suppression of the super-Hubble $\chi$ fluctuation. 
In addition to this, although we did not consider the interaction 
between $\sigma$ and $\chi$ fields, including this coupling may 
strengthen parametric resonance of scalar fields\cite{Bruce}. 
It is quite interesting to investigate the evolution of metric and 
field fluctuations in broad classes of models
with several interactions in the sense that 
we can constrain the inflaton potential 
in terms of distortions from the CMB spectrum.

Although we have investigated compactifications
on the sphere, there exist several ways of
compactifications on other manifolds.
For example, the dilaton does not have its own potential 
in the torus compactification, which would lead to 
the growth of extra dimensions during inflation as was
analyzed in Ref.~\cite{BM}.
In this case, since the resonance term 
$g^2\phi^2e^{-d\sigma/\sigma_*}$ in the l.h.s.
of Eq.~$(\ref{C7})$ will be suppressed during inflation,
the ordinary picture of preheating may be modified in the 
presence of extra dimensions.
It is also of interest to investigate the dynamics of 
inflation and preheating in more realistic models
of compactifications such as Calabi-Yau manifolds,
because it is possible to judge whether such 
compactifications are appropriate or not from 
the cosmological point of view.
These issues are under consideration.

\section*{ACKNOWLEDGMENTS}
The author would like to thank Bruce A. Bassett, Kei-ichi Maeda, 
Shinji Mukohyama, Takashi Torii, Kunihito Uzawa, Fermin Viniegra,
and Hiroki Yajima for useful discussions and comments. 
This work was supported partially by a Grant-in-Aid for  Scientific
Research Fund of the Ministry of Education, Science and Culture
(No. 09410217), and by the Waseda University 
Grant for Special Research Projects.


\end{document}